\documentclass[intlimits,twoside,a4paper]{article}
\usepackage[cp1251]{inputenc}
\usepackage{graphicx}
\usepackage{wrapfig}
\usepackage[T2A]{fontenc}

\usepackage{cmpj3}

\issue{2023}{26}{4}{43701}
\doinumber{10.5488/CMP.26.43701}

\title[Dynamic behavior of the antiferromagnetically coupled bilayer Ising model]{Dynamic behavior of the antiferromagnetically coupled bilayer Ising model}

\author[C. Ekiz, R. Erdem, D. Semet]{C. Ekiz\orcid{0000-0001-8943-5235}\refaddr{label1}, R. Erdem\refaddr{label2}, D. Semet\refaddr{label3}}

\addresses{
\addr{label1} Department of Physics, Faculty of Science, Ayd\i n Adnan Menderes University, 09010 Ayd\i n, T\"urkiye \\ 
\addr{label2} Department of Physics, Faculty of Science, Akdeniz University, 07058 Antalya, T\"urkiye \\
\addr{label3} Research and Application Hospital, Ayd\i n Adnan Menderes University, 09010 Ayd\i n, T\"urkiye \\
}

 	
\date{Received May 2, 2023, in final form August 30, 2023}
\begin{document}

\maketitle

\begin{abstract}
Using the path probability and lowest approximation of cluster variation method, we study the dynamic and equilibrium properties of a bilayer magnetic system, consisting of two ferromagnetic monolayers  antiferromagnetically coupled for different spins $(\sigma=1/2$ and $S=1)$. Firstly, numerical results of the monolayer and total magnetizations are presented under the effect of the diverse physical parameters, and the phase diagrams of bilayer system are discussed. Then, since it is well established that the path probability method is an effective method for the existence of metastable states, the time evolution of monolayer- and total magnetizations is investigated.  
\printkeywords
\end{abstract}

\section{Introduction}
Ferrimagnets have two or more sublattices that are magnetized in different directions. Ferrimagnetic materials remain magnetic until a critical point called N\'eel temperature is reached, where thermal motion randomizes the magnetic spins. Spinel ferrites exhibit the most prevalent ferrimagnetic behavior, which is characterized by adjacent antiparallel electron spins with different magnitudes. The possibility that the net moment can change the sign at a compensation temperature below the N\'eel temperature is intriguing, and they are important for microelectronic and magneto-optical recording  \cite{manriquez}. The N\'eel and compensation temperatures are different for each material and span a wide range. Since the magnetic properties of layered composite materials may differ from their corresponding bulk systems, they have received considerable attention. The mixed and pure layered Ising systems with antiferromagnetic exchange interaction provide simple but interesting models for studying layered composite materials \cite{bengrine,du1,du2,ekiz2003,ekiz2004,saber,jiang1,wiatrowski,jiang2,espriella,jabar1,jabar2,erta,rosin}. 

The physics of first-order phase transitions is where the idea of metastability initially appeared. From one atom to statistical ensembles of molecules, such as binary fluids, binary alloys, amorphous solids, liquid crystals, minerals, magnetic systems at molecular levels or as a whole, metastability is a common state in physics and chemistry. The quantity of states increases with system size and in case the factors driving their reciprocal interaction are less uniform or more diversified in space. The time-invariance of the active or reactive patterns with respect to the external effects defines stability and metastability in dynamic systems including magnetic systems \cite{ekiz02,erdem,ekiz99,santos}, electrical circuits, signal trafficking, decisional, neurological and immunological systems.

 Investigations of the critical phenomena for the bilayer Ising systems are needed for a basic scientific knowledge on the magnetic properties. The magnetic properties of a ferrimagnetic bilayer system consisting of ferromagnetically coupled monolayers (A and B) with mixed spins (1/2 and 1) were investigated using the exact recursion method \cite{ekiz08}. It was shown that a mixed spin ferrimagnetic bilayer system may have one tricritical point and compensation points on the recursive lattice in agreement with the previous studies. In recent times, Jiang et al. \cite{su}  investigated a bilayer nano-stanene-like structure with Ruderman-Kittel-Kasuya-Yosida (RKKY) coupling described by the Ising model. They found that the exchange coupling, external magnetic field, the number of nonmagnetic layers and the anisotropy had major effects on the magnetization process, specific heat and internal energy. In another study, the magnetic properties, magnetocaloric effect and hysteresis behaviors were investigated and predicted for a mixed-spin AFM/FM Ising BiFe$_3$/Co bilayer applying the Monte-Carlo simulation~\cite{chang}. The magnetization, susceptibility and critical temperature were investigated under various exchange couplings and external magnetic fields. The simulation results indicated that a decrease of the exchange coupling and an increase of external magnetic fields can cause an increase of magnetic entropy change, adiabatic temperature change and relative cooling power.  In addition to the equilibrium properties using several methods \cite{si,fadil,bahmad,masrour,li,wang,harir,diaz,xu,badarneh}, it is important to examine the dynamic properties of the layered mixed system. In this way, by examining the time dependent variations of the monolayer magnetizations, the existence of metastable states as well as stable state can be examined. Thus, our main motivation in this study is to examine the dynamic properties of the bilayer system, whose magnetic properties in equilibrium state on the Bethe lattice were examined before, using the path probability method (PPM). The agreement of the results obtained from PPM will also be compared with the finite temperature results obtained from the lowest approximation of cluster variation method (LACVM).

This article is organized as follows. The bilayer Ising model and examination techniques are introduced in section \ref{model}. In section \ref{result}, the dynamics of monolayer and total magnetizations, free energy, compensation behavior and phase diagrams are discussed. The conclusions of this study are briefly presented in section~\ref{conclusion}.

\section{Bilayer Ising model and calculation methods}
\label{model}

\begin{figure}[!b]
\begin{center}
\includegraphics[width=0.5\textwidth]{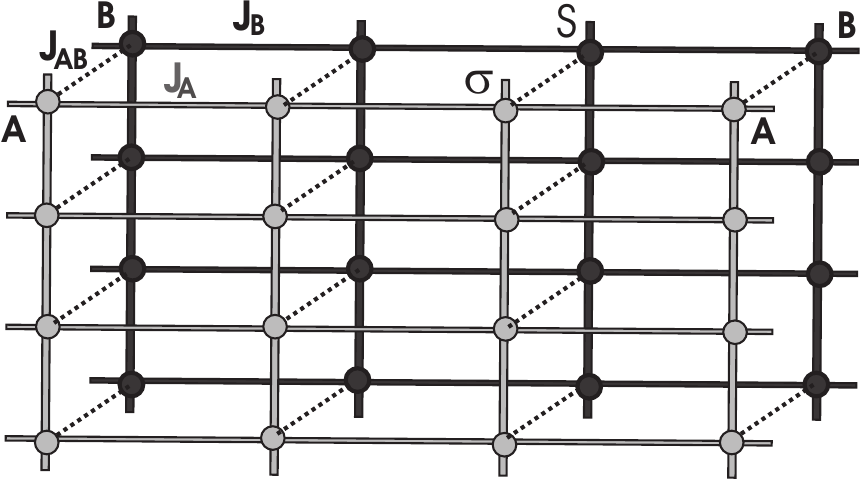}
\end{center}
\caption{Part of the bilayer system constructed of two ferromagnetic monolayers (A and B) coupled with an interlayer interaction parameter $J_{AB}$.}
\label{fig1}      
\end{figure}
We consider a bilayer Ising model consisting of two magnetic monolayers (A and B) with different spins, $\sigma$   and $S$. The model is described as a two-layer system seen in figure \ref{fig1} with spin variables $\sigma_{i}=\pm 1/2$ and $S_{i'}=\pm 1,0$ on the magnetic monolayers A and B, respectively. In this way, the Hamiltonian of a two-layer Ising model is given by 
\begin{eqnarray}
{\cal H} = -J_{A}\sum_{\langle i,j \rangle }^{Nq/2} \sigma_{i} \sigma_{j}-J_{B}\sum_{\langle i',j' \rangle }^{Nq/2} S_{i'} S_{j'}-J_{AB}\sum_{\langle i,i' \rangle }^{N}\sigma_{i} S_{i'}-D\sum_{\langle i' \rangle }^{N}(S_{i'})^2.
\label{eq:1}
\end{eqnarray}
The first two summations in the model Hamiltonian are taken over the nearest neighboring interacting spin pairs on the same monolayers. For the ferromagnetic interactions on each monolayer, the bilinear interaction parameters are considered as $J_{A}>0$, $J_{B}>0$ and if the spin interactions between adjacent layers are antiferromagnetic, they are taken as  $J_{AB}<0$. Furthermore, $N$ is total numbers of atoms, and $q$ is the coordination number of the lattice. Finally, parameter $D$ measures the strength of uniaxial single-ion anisotropy acting on only B monolayer with spin-$S$ sites.
For the spin system at equilibrium, we can derive the set of self-consistent equations for the order parameters using the lowest approximation of the cluster variation method (LACVM) \cite{kikuchi51,tanaka,ekiz02}. In this method, the average value of each of the spin states can be represented by $x_{1}^{A}$, $x_{2}^{A}$  on the sites of monolayer A and $x_{1}^{B}$, $x_{2}^{B}$ and $x_{3}^{B}$ on the monolayer B, which are called the state variables. $x_{1}^{A}$ is the fraction of spin value 1/2 on monolayer A and $x_{2}^{A}$ is the fraction of the spin that has the value  $-1/2$ on the same monolayer. Similarly, $x_{1}^{B}$ is the fraction of spin value $+1$, $x_{2}^{B}$ is the fraction of spin value 0 and $x_{3}^{B}$ is the fraction of spin value $-1$ on monolayer B. The above variables satisfy the normalization conditions for monolayers A and B, respectively:
\begin{eqnarray}
\sum_{i=1}^{2}x_{i}^A=1, \quad
\sum_{i'=1}^{3}x_{i'}^B=1.
\label{eq:2}
\end{eqnarray}
The problem here is the evaluation of the spin mean values. The order parameters in terms of internal variables are given by 
\begin{eqnarray}
m_{A}=\left \langle \sigma_{i} \right \rangle=\frac{1}{2}(x_{1}^A-x_{2}^A), \quad
m_{B}=\left \langle S_{i'} \right \rangle=x_{1}^B-x_{3}^B, \quad
q_{B}=\left \langle (S_{i'})^2 \right \rangle=x_{1}^B+x_{3}^B.
\label{eq:3}
\end{eqnarray}
Using (\ref{eq:2})--(\ref{eq:3}), the internal variables can be calculated as linear combinations of the order parameters
\begin{eqnarray}
x_{1}^A=\frac{1}{2}(1+2m_{A}), \qquad
x_{2}^A=\frac{1}{2}(1-2m_{A}),
\label{eq:4}
\end{eqnarray}

\begin{eqnarray}
x_{1}^B=\frac{1}{2}(m_{B}+q_{B}), \quad
x_{2}^B=(1-q_{B}), \quad
x_{3}^B=\frac{1}{2}(q_{B}-m_{B}).
\label{eq:5}
\end{eqnarray}
In order to study the equilibrium behaviors of the model, it is necessary to calculate the magnetizations and the quadrupolar moment for each monolayer A and B. In the framework of LACVM \cite{kikuchi51,tanaka,ekiz02}, which is identical to the mean-field approximation \cite{strecka15}, a simple expression of the internal energy of the model gives the following expression for a 2D square lattice monolayer shown in  figure \ref{fig1}:
\begin{eqnarray}
\frac{E}{N}=-2J_{A}m_{A}^2-2J_{B}m_{B}^2-J_{AB}m_{A}m_{B}-Dq_{B}.
\label{eq:6}
\end{eqnarray}
If we substitute (\ref{eq:4})--(\ref{eq:5}) into (\ref{eq:7}), the internal energy per lattice sites can be written as:
\begin{eqnarray}
\frac{E}{N}=-\frac{1}{2}J_{A}(x_{1}^A-x_{2}^A)^2-2J_{B}(x_{1}^B-x_{3}^B)^2-\frac{1}{2}J_{AB}(x_{1}^A-x_{2}^A)(x_{1}^B-x_{3}^B)-D(x_{1}^B+x_{3}^B).
\label{eq:7}
\end{eqnarray}
The total entropy $S$ and free energy $F$ are respectively given by 
\begin{eqnarray}
S=k_{\text B} \ln \Omega, \quad
F=E-TS,
\label{eq:8}
\end{eqnarray}
where $\Omega$ is the weight factor. In this approximation, the monolayer weight factors $\Omega^A$ and $\Omega^B$ can be expressed in terms of internal variables for monolayers A and B, respectively, as follows:
\begin{eqnarray}
\Omega^A=\frac{(N!)}{\prod_{i=1}^{2}(x_{i}^AN)!}, \quad
\Omega^B=\frac{(N!)}{\prod_{i'=1}^{3}(x_{i'}^B N)!}. 
\label{eq:9}
\end{eqnarray}
Using equations (\ref{eq:7})--(\ref{eq:9}) and making use of the Stirling formula, the free energy for per site ($\phi=-\beta F/N$) can be now written as 
\begin{eqnarray}
	\phi &=&-\frac{\beta F}{N}=\frac{1}{2}\beta J_{A}(x_{1}^A-x_{2}^A)^2+2\beta J_{B}(x_{1}^B-x_{2}^B)^2+\frac{1}{2}\beta J_{AB}(x_{1}^A-x_{2}^A)(x_{1}^B-x_{3}^B)+
\nonumber \\
&& \beta D(x_{1}^B+x_{3}^B)-\Bigl \{ \sum_{i=1}^{2}x_{i}^{A}\ln x_{i}^{A}+\sum_{i'=1}^{3}x_{i'}^{B}\ln x_{i'}^{B} \Bigr \}+\lambda\Bigl ( 1-\sum_{i=1}^{2} x_{i}^A\Bigr )+
\lambda\Bigl ( 1-\sum_{i'=1}^{3} x_{i'}^B\Bigr ),
\label{eq:10}
\end{eqnarray}
where $\lambda$ is presented for obtaining the normalization requirement. $\beta=1/k_{\text{B}}T$, $T$ is the absolute temperature and $k_{\text{B}}$ is the Boltzmann constant. Thus, the self-consistent equations for the three long-range order parameters, namely $m_{A}$, $m_{B}$ and $q_{B}$ are derived by using the following minimization formulae:
\begin{eqnarray}
\frac{\partial \phi}{\partial x_{i}^A}=0, \quad	
\frac{\partial \phi}{\partial x_{i'}^B}=0.
\label{eq:11}
\end{eqnarray}
Using equations (\ref{eq:10})--(\ref{eq:11}), the self-consistent equations or monolayer magnetizations and quadrupolar moment are found to be
\begin{eqnarray}
m_{A}\!\!\!&=&\!\!\!\frac{1}{2}\tanh\left [ \beta (2J_{A}m_{A}+\frac{1}{2}J_{AB}m_{B}) \right ],
\nonumber \\
m_{B}\!\!\!&=&\!\!\!\frac{2\exp{(\beta D)}\sinh\left [ \beta (4J_{B}m_{B}+J_{AB}m_{A}) \right ]}{1+2\exp{(\beta D)}\cosh\left [ \beta (4J_{B}m_{B}+J_{AB}m_{A}) \right ]}, \nonumber \\
q_{B}\!\!\!&=&\!\!\!\frac{2\exp{(\beta D)}\cosh\left [ \beta (4J_{B}m_{B}+J_{AB}m_{A}) \right ]}{1+2\exp{(\beta D)}\cosh\left [ \beta (4J_{B}m_{B}+J_{AB}m_{A}) \right ]}.
\label{eq:12}
\end{eqnarray}
By solving the nonlinear algebraic equations (\ref{eq:12}) numerically, one can show the thermal variations of the order parameters $m_{A}$, $m_{B}$ and $q_{B}$ for various values of $J_{B}/J_{A}$, $D/J_{A}$ and $J_{AB}/J_{A}$. Examples of the solutions will be given and discussed in the next section.

For a system away from equilibrium, the general time dependence of state variables is given by the PPM \cite{kikuchi} as,
\begin{eqnarray}
\frac{\rd x_{i}}{\rd t}=\sum_{i\neq j} \left ( \chi_{ji}-\chi_{ij} \right ),
\label{eq:13}
\end{eqnarray} 
where $\chi_{ji}$ is the path probability rate for the system to go from state $i$ to $j$. Using this technique, we may examine relaxation curves of the order parameters and observe the flatness property of metastable states. The PPM has been successfully applied to describe the dynamic behavior of many homogeneous and non-homogeneous stationary systems.
The coefficients $\chi_{ij}$  are the product of three factors: $k_{ij}$  the rate constants with $k_{ij}= k_{ji}$, a temperature-dependent factor which ensures that the time-independent state is the equilibrium state and a third factor which is the fraction of the system that is in the state $i$. A detailed equilibration requires $\chi_{ij}=\chi_{ji}$. To determine the path probability rate, the following formula is generally employed
\begin{eqnarray}
\chi_{ij}=\frac{k_{ij}}{Z}\exp\left ( -\beta\frac{\partial E}{\partial x_{j}} \right ),
\label{eq:14}
\end{eqnarray} 
where $E$ is the internal energy presented in equation (\ref{eq:8}). Using equation (\ref{eq:3}), the time dependent rate equations are written as follows: 
\begin{eqnarray}
\tfrac{\rd m_{A}}{\rd t}\!\!\!&=&\!\!\!\tfrac{1}{2}\left ( \tfrac{\rd x_{1}^A}{\rd t}-\tfrac{\rd x_{2}^A}{\rd t} \right ),\nonumber\\
\tfrac{\rd m_{B}}{\rd t}\!\!\!&=&\!\!\!\left ( \tfrac{\rd x_{1}^B}{\rd t}-\tfrac{\rd x_{3}^B}{\rd t} \right ),\nonumber\\
\tfrac{\rd q_{B}}{\mathrm{\rd} t}\!\!\!&=&\!\!\!\left ( \tfrac{\rd x_{1}^B}{\rd t}+\tfrac{\rd x_{3}^B}{\rd t} \right ).
\label{eq:15}
\end{eqnarray}
Using equations(\ref{eq:3})--(\ref{eq:7}), (\ref{eq:13})--(\ref{eq:15}), the time dependent dynamic equation for monolayer magnetization $m_{A}$ is obtained as follows:
\begin{eqnarray}
\tfrac{Z_{A}}{k_{1}}\tfrac{\rd m_{A}}{\rd t}=\left(\tfrac{1}{2}+m_{A}\right) \xi _{2}^A-\left(\tfrac{1}{2}-m_{A}\right)\xi_{1}^A.
\label{eq:16}
\end{eqnarray} 
Then, the same procedure above is applied to the monolayer B for the general Hamiltonian given in~(\ref{eq:1}), the following two dynamic equations are derived for magnetization monolayer B and quadrupolar moment, respectively:
\begin{eqnarray}
\frac{Z_{B}}{k_{1}}\frac{\rd m_{B}}{\rd t}\!\!\!&=&\!\!\!(\xi_{1}^B-\xi_{3}^B)+(k-1)(\xi_{1}^B-\xi_{3}^B)q_{B}-(k\xi_{1}^B+\xi_{2}^B +k\xi_{3}^B)m_{B}, \nonumber\\
\frac{Z_{B}}{k_{1}}\frac{\rd q_{B}}{\rd t}\!\!\!&=&\!\!\!(\xi_{1}^B+\xi_{3}^B)-(\xi_{1}^B+\xi_{2}^B +\xi_{3}^B)q_{B},
\label{eq:17}
\end{eqnarray} 
where $k=k_{2}/k_{1}$, while the exponential expressions ($\xi_{i}^A$ and $\xi_{i'}^B$) and partition functions of monolayers ($Z_{A}$ and $Z_{B}$) are calculated from the following general expressions respectively, as:
\begin{eqnarray}
\xi_{i}&=&\exp(-\beta\frac{\partial E}{\partial x_{i}}),\\
Z&=&\sum_{i} \xi_{i}.
\label{eq:18}
\end{eqnarray} 
After deriving the dynamical state equations, we can determine their solutions to see the non-equilibrium properties and discuss regarding the results obtained for finite temperature behavior. To this end, (\ref{eq:16})--(\ref{eq:17}) are solved by the fourth-order Runge-Kutta method \cite{butcher,press}. The Runge-Kutta method is a widely used and effective tool to solve the initial value problems of differential equations. It can be used to build precise numeric methods of high-order by functions without the need for higher order derivatives of expressions.

\section{Numerical results and discussion}\label{result}
\begin{figure}[!b]
\centering	{\includegraphics[width=0.45\textwidth]{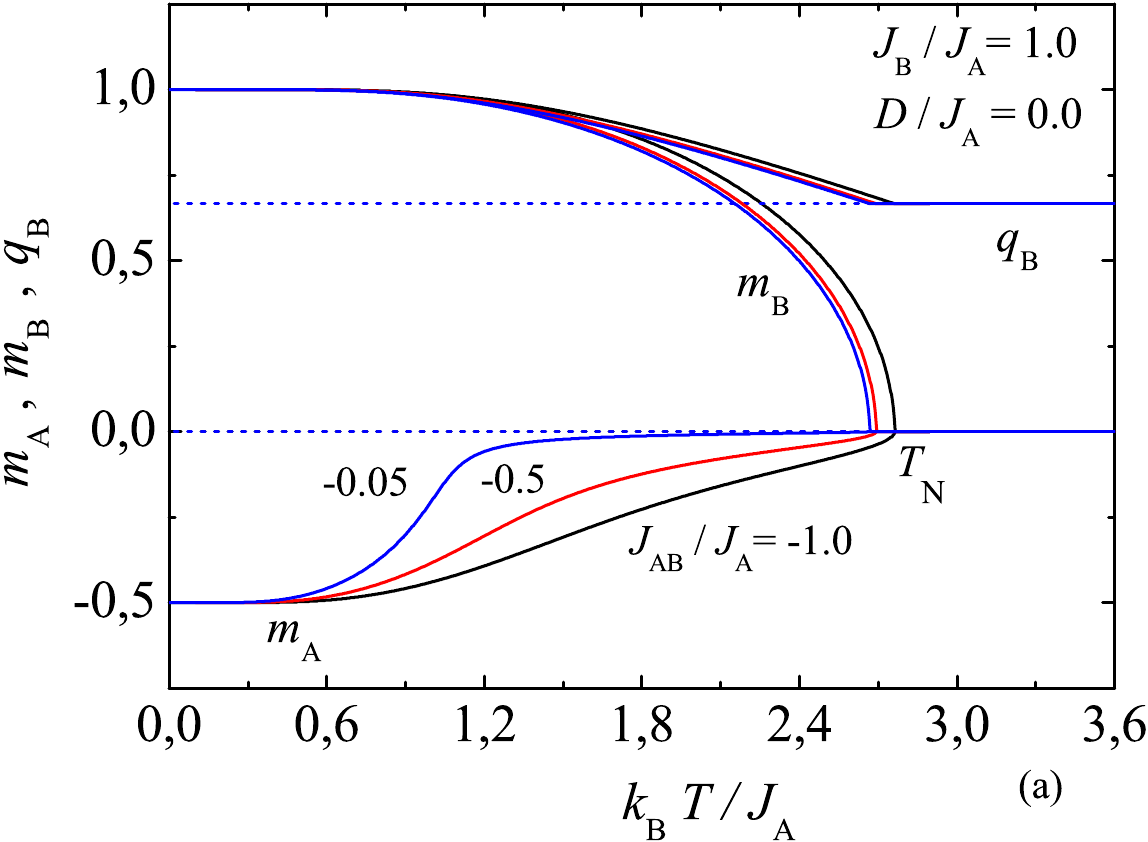}
	\includegraphics[width=0.45\textwidth]{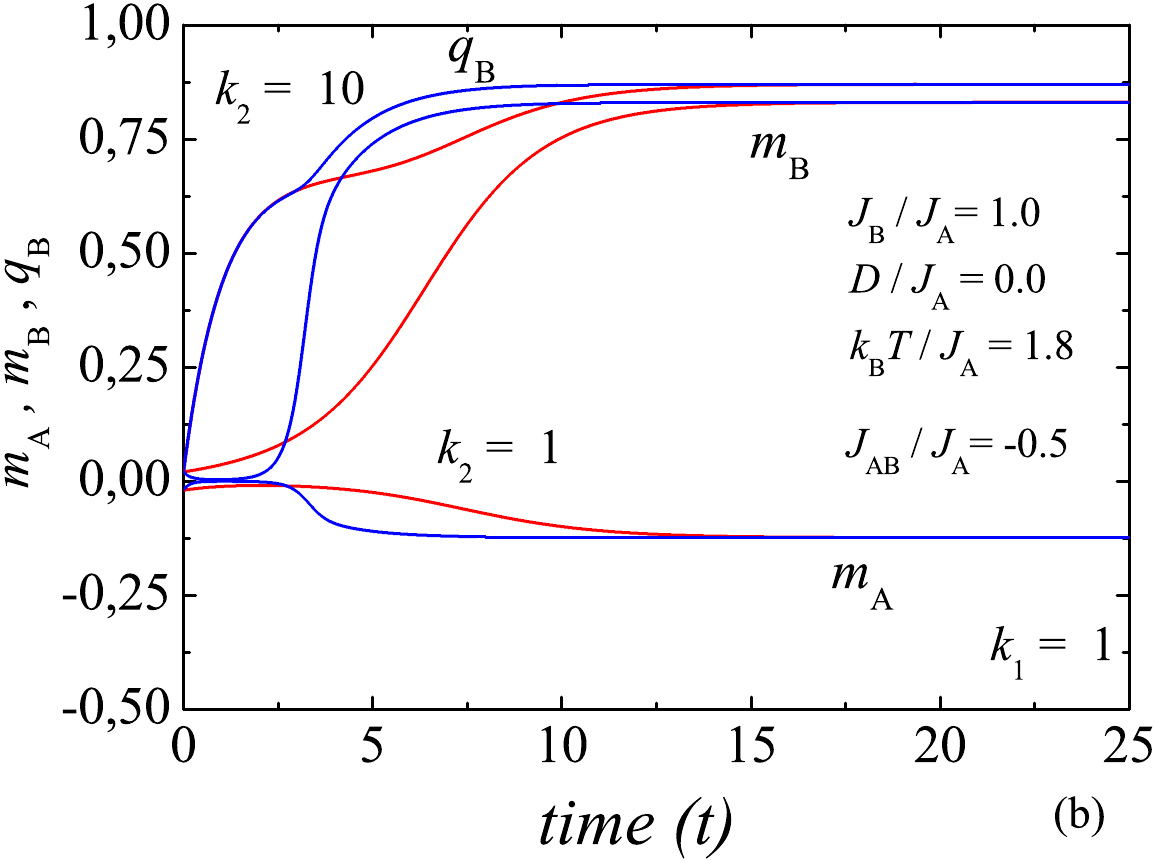}
	\includegraphics[width=0.45\textwidth]{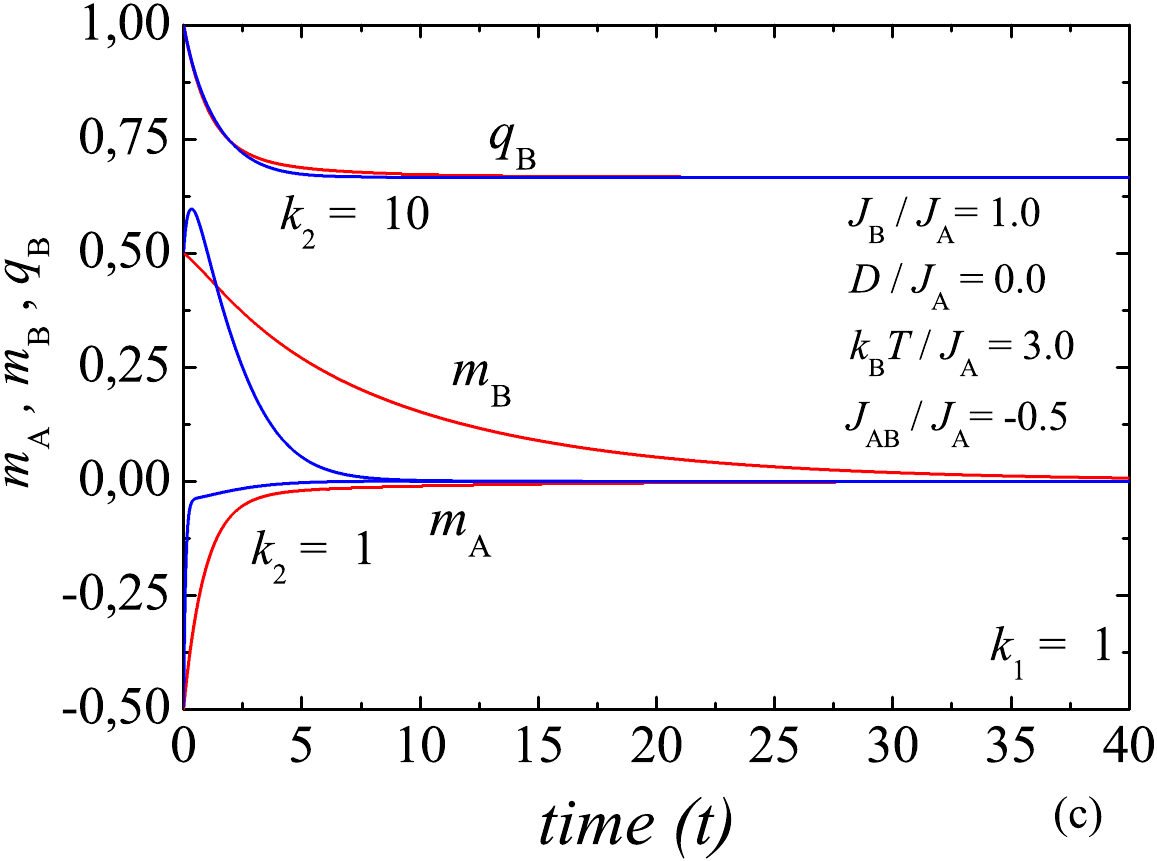}}
	\caption{(Colour online) (a) The temperature dependence of monolayer magnetizations $m_{A}$, $m_{B}$ and quadrupolar moment $q_{B}$ for selected interaction parameters in the absence of single-ion anisotropy.
		(b) The time evolution of $m_{A}$, $m_{B}$ and $q_{B}$ at ferrimagnetic phase for $k_{\text{B}}T/J_{A}=1.8$  and  $k_{1}=1; k_{2}=1,10$.
		(c) The time evolution of $m_{A}$, $m_{B}$ and $q_{B}$ at paramagnetic phase for $k_{\text{B}}T/J_{A}=3.0$  and  $k_{1}=1; k_{2}=1,10$.}
	\label{fig2}       
\end{figure}
In this section, we  consider the antiferromagnetic interaction between monolayers in all our numerical calculations. When the interaction parameter between A and B monolayers is selected as the antiferromagnetic, the monolayer magnetizations exhibit antisymmetric orientation at ground state and finite temperatures. The variation of monolayer magnetizations with temperature can be easily obtained numerically from the equations (\ref{eq:12}). The monolayer  magnetizations and quadrupolar moment derived from the dynamical equations of the model are identical to the equations obtained by minimizing the free energy with respect to the internal variables in  equation (\ref{eq:11}). In the bilayer Ising model, the different spin magnitudes and interaction parameters at the monolayers significantly affect the magnetic and dynamic properties of the bilayer model.

First, let us begin by examining the variation of magnetizations with temperature to analyze the time dependence of monolayer magnetizations and quadrupolar moment. By obtaining the equilibrium state relations of monolayer magnetizations in the time-independent state of the dynamical equations $(\frac{\mathrm{d} m_{A}}{\mathrm{d} t}=\frac{\mathrm{d} m_{B}}{\mathrm{d} t}=\frac{\mathrm{d} q_{B}}{\mathrm{d} t}=0)$, their thermal variations can also be investigated. In the absence of the single-ion anisotropy parameter $D/J_{A}$, the thermal variations of the monolayer magnetizations and quadrupolar moment can be seen in figure \ref{fig2}(a). In this case, the symmetry breaking is evident due to the different ground state spin orientations and the spin interactions between the monolayers. Thus, the bilayer Ising model undergoes a continuous phase transition at N\'eel temperature $T_{N}$. In figure~\ref{fig2}(b), the time-dependent variations, i.e., the dynamics of monolayer magnetizations and quadrupolar moment are given for the same parameters as in figure \ref{fig2}(a). For arbitrarily chosen initial values at the time  $t=0$, monolayer magnetizations and quadrupole moments relax after sufficient time and become time-independent at a certain fixed temperature as time progresses. If  figure \ref{fig2}(a) and figure \ref{fig2}(b) are compared at the temperature $k_{\text{B}}T/J_{A}=1.8$, one can  see that the equilibrium value of the monolayer magnetizations and the time-dependent results obtained from the relaxation curves using the dynamic equations agree exactly. The arbitrary initial values of the monolayer magnetizations were chosen as $m_{A}=-0.02$, $m_{B}=0.02$  and $q_{B}=0.02$ for monolayers A and B, respectively. In any case, if only stable states with the lowest free energy exist, the time evolution for all values of monolayer magnetizations will eventually be at the fixed equilibrium states of $m_{A}=-0.123$, $m_{B}=0.83$  and $q_{B}=0.87$. On the other hand, if there are some metastable states with local minimum free energy in the bilayer magnetic system at this fixed temperature, the order parameters should also relax at the metastable states, considering all possible initial values. The classification of these states can be done by methods such as free energy analysis, solving dynamic equations and flow diagrams. Since we have seen that the system always relaxes into just one state, the metastable state does not exist in present case. When the time dependence of the layered magnetization is calculated for various $k$ values, it will be seen that there is an inverse relationship between the relaxation time $\tau$ and  parameter $k$. When $k_{2}>k_{1}$, for the same initial values of monolayer magnetizations and quadrupolar moments, the bilayer system relaxes into the stable state faster than $k_{1}=k_{2}$. Here, the first-rate constants are $k_{13}=k_{32}=k_{1}$ which is the insertion or removal of particles associated with the translation of particles through the lattice. The second-rate constant $k_{12}=k_{2}$ is associated with the reorientation of a particle at a fixed site. More specifically, the relaxation time takes a long time for the order parameters in the case of $k=1$, while the same case gets a shorter time period with increasing values of $k_{2}$. Figure \ref{fig2}(c) denotes the dynamic relaxation curves of the order parameters at a fixed temperature $k_{\text{B}}T/J_{A}=3.0$ in the paramagnetic phase. Since there are only stable states at the selected reduced temperature, there is a relaxation at $m_{A}=m_{B}=0$ and $q_{B}=0.666$ for all initial values. In addition, it is seen that the relaxation time shortens with the increase of $k_{2}$.
\begin{figure}[!t]
	\includegraphics[width=0.46\textwidth]{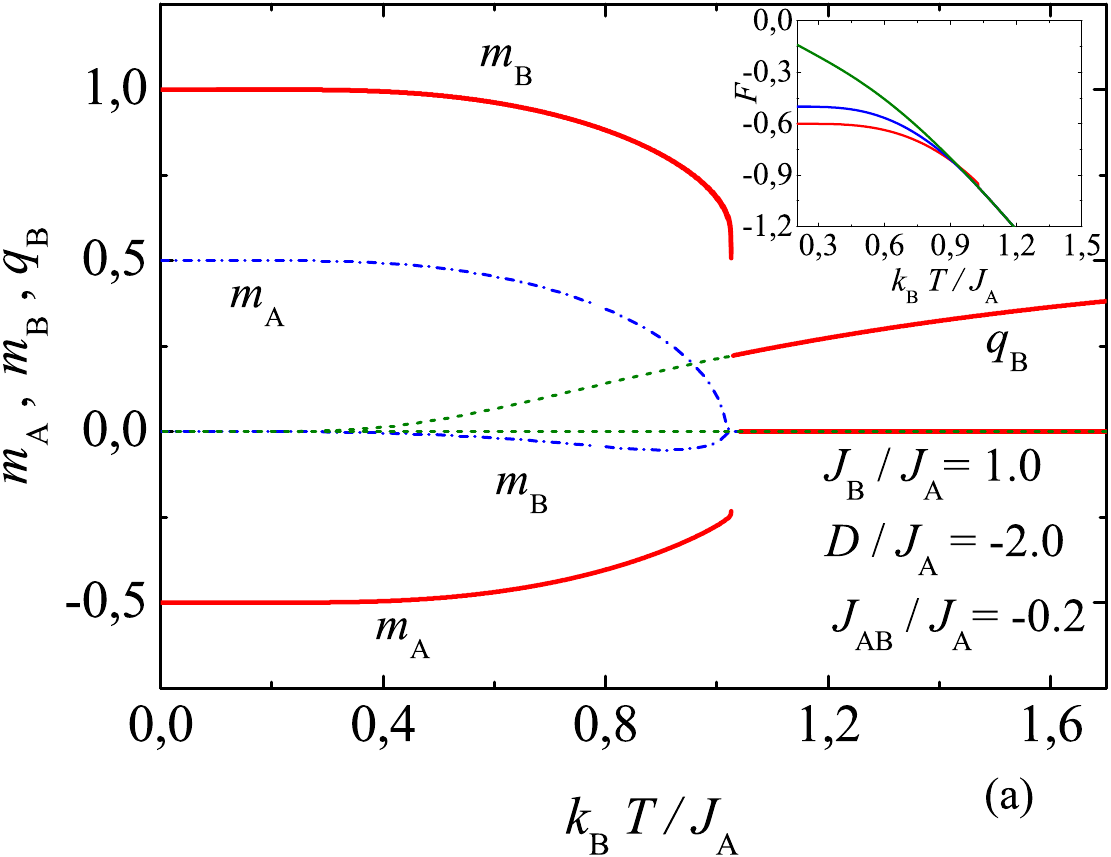}
	\includegraphics[width=0.49\textwidth]{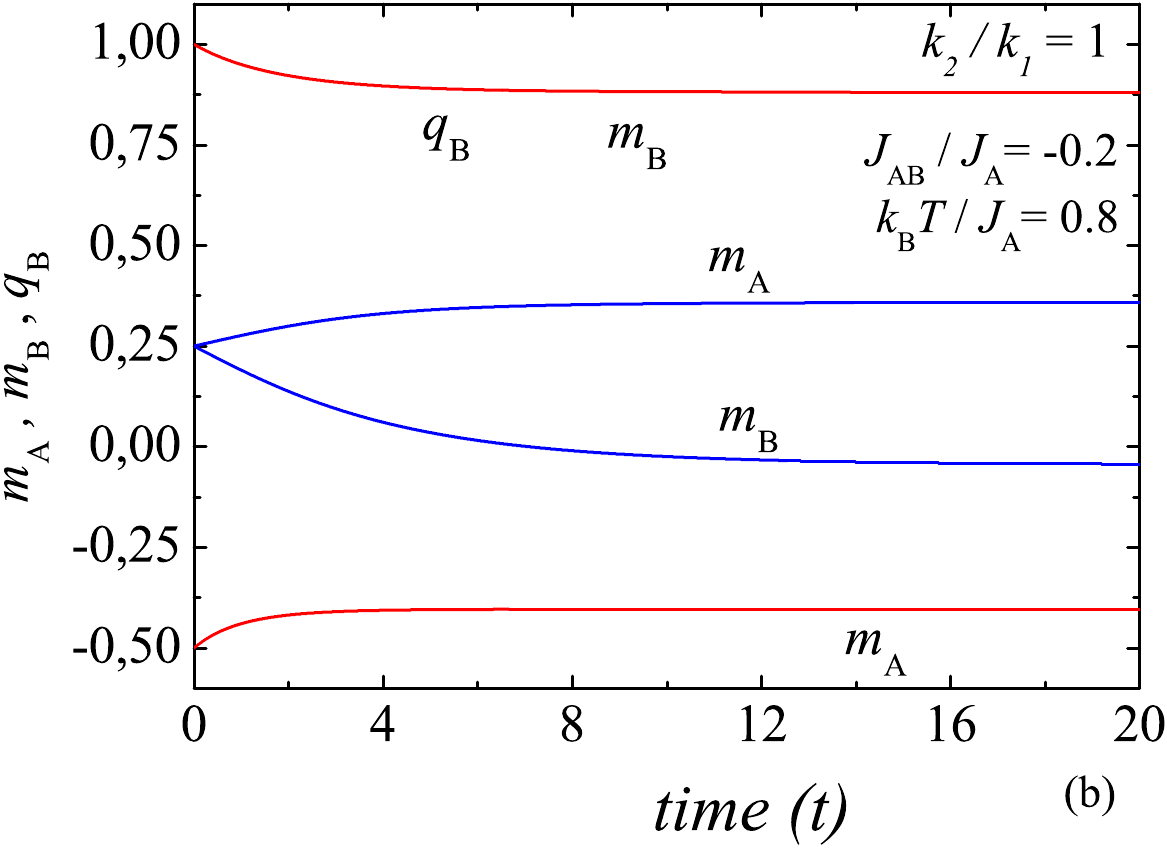}
	\caption{(Colour online) (a) The temperature dependence of $m_{A}$, $m_{B}$ and $q_{B}$ for the selected interaction parameters and single-ion anisotropy.
		(b) Relaxation curves of $m_{A}$, $m_{B}$ and $q_{B}$ for $J_{AB}/J_{A}=-0.2$, $D/J_{A}=-2.0$ and  $k_{\text{B}}T/J_{A}=0.8$.}
	\label{fig3}       
\end{figure}

It is known that the interaction parameters in the model Hamiltonian have a significant effect on the equilibrium and dynamic properties. For example, with inclusion of the single-ion anisotropy parameter for the monolayer B, a first-order phase transition is seen for both monolayer order parameters in figure \ref{fig3}(a). In the bilayer Ising model considered, the first-order phase transition is not observed for $J_{AB}/J_{A}=-1$, but this behavior is observed at sufficiently large values such as $J_{AB}/J_{A}=-0.2$ and $-0.05$. Below the first-order phase transition temperature of the monolayer magnetizations, the metastable states should be expected because the metastability is closely related to the first-order phase transitions. Using an appropriate numerical analysis method, the stable, metastable and unstable states of the monolayer order parameters at any temperature can be obtained and classified. In the figure, solid lines indicate the stable states, dashed-dotted lines correspond to the metastable states, and dashed lines also show unstable states. This classification is done by comparing the free energy values of these solutions. The relaxation curves of the monolayer order parameters obtained as a result of dynamic equation solution are shown in figure \ref{fig3}(b). In the solution of dynamic equations, the temperature $k_{\text{B}}T/J_{A}=0.8$ is considered for arbitrary initial values. Since unstable states with the highest free energy cannot be determined from the relaxation curves, it is seen that they are stable and metastable states. Using the free energy analysis seen in the inset  in figure \ref{fig3}(a), it is determined that the solid lines are stable and the dashed-dotted lines\newpage \noindent  are metastable states.
\begin{figure}[!t]
\includegraphics[width=0.48\textwidth]{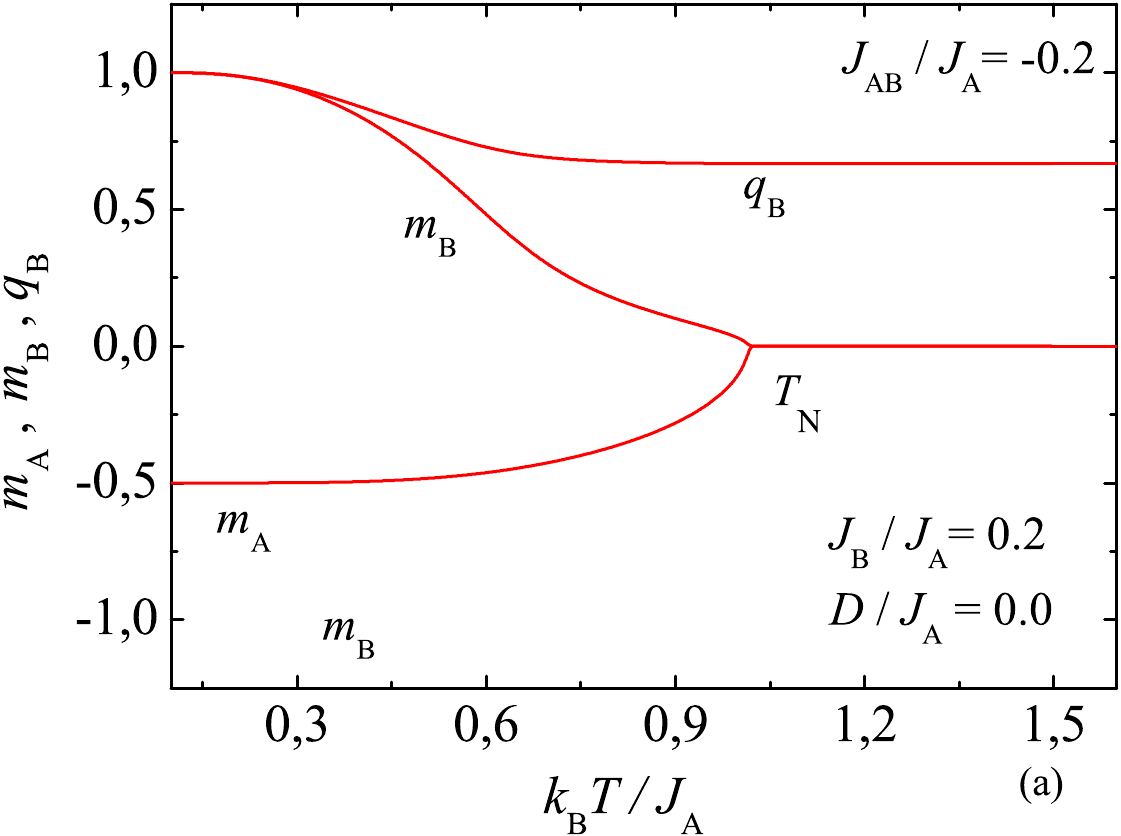}
\includegraphics[width=0.48\textwidth]{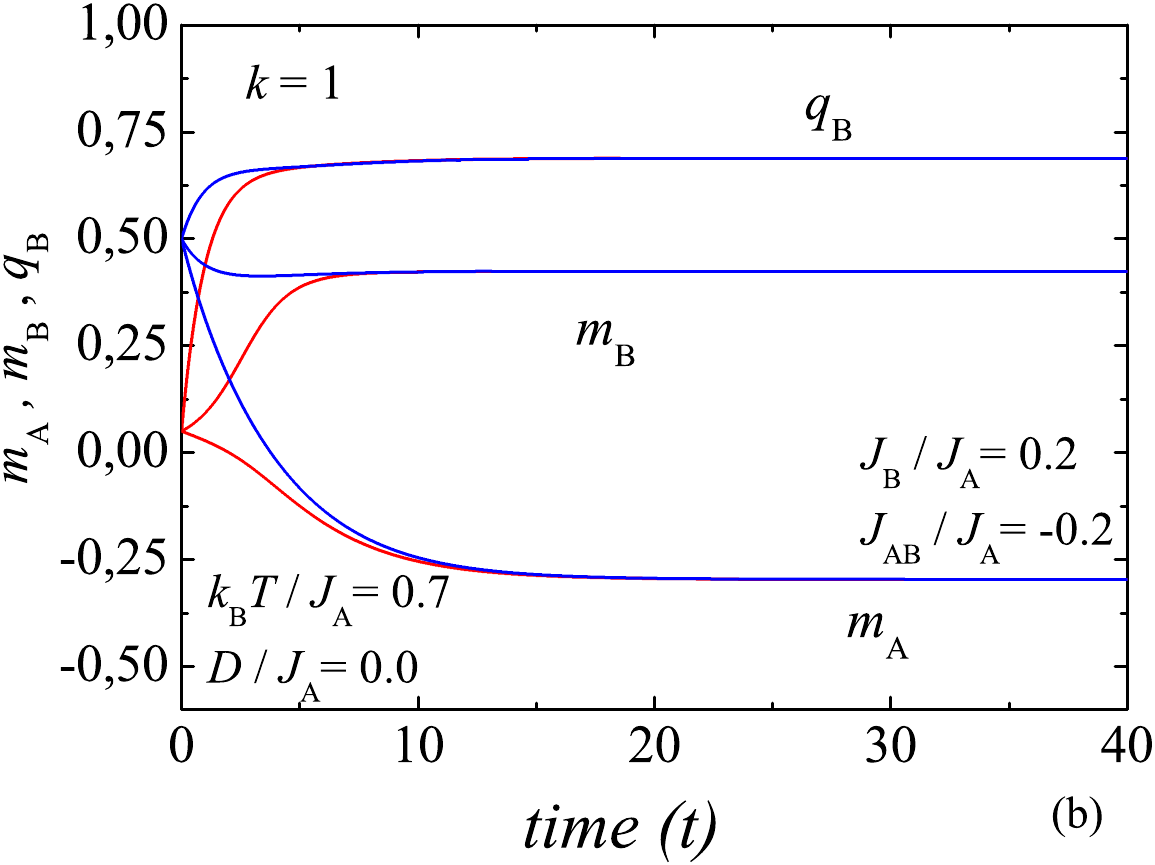}
\caption{(Colour online) (a) The temperature dependence of $m_{A}$, $m_{B}$ and $q_{B}$ for selected interaction parameters and single-ion anisotropy. (b) Relaxation curves of $m_{A}$, $m_{B}$ and $q_{B}$ for $J_{AB}/J_{A}=-0.2$, $D/J_{A}=0$  and  $k_{\text{B}}T/J_{A}=0.7$.}
\label{fig4}       
\end{figure}
It should be noted here that if the interaction parameters of monolayers are different from each other, that is, when $J_{A}\neq J_{B}$, the thermal variation of magnetizations and quadrupolar moment will be different. This situation is clearly seen for $J_{B}/J_{A}=0.2$ in figure \ref{fig4}(a). In figure \ref{fig4}(b), the dynamic behavior of the order parameters for $J_{AB}/J_{A} =-0.2$, $J_{B}/J_{A}=0.2$ and $k_{\text{B}}T/J_{A}=0.7$ indicates that the possible time evolution of the order parameters is compatible with the stable states observed from the finite temperature behavior seen in figure \ref{fig4}(a). When the temperature is less than the second-order phase transition temperature,
the system always comes to a stable state. Therefore, relaxation processes are not influenced by the rate constants and initial values of order parameters. It should be mentioned that when the temperature reaches the continuous phase transition, the system requires an excessively lengthy period to relax into disorder states. On the other hand, if the temperature exceeds critical temperature, it takes short time for the system to relax into disorder states. 

\begin{figure}[!t]
\includegraphics[width=0.48\textwidth]{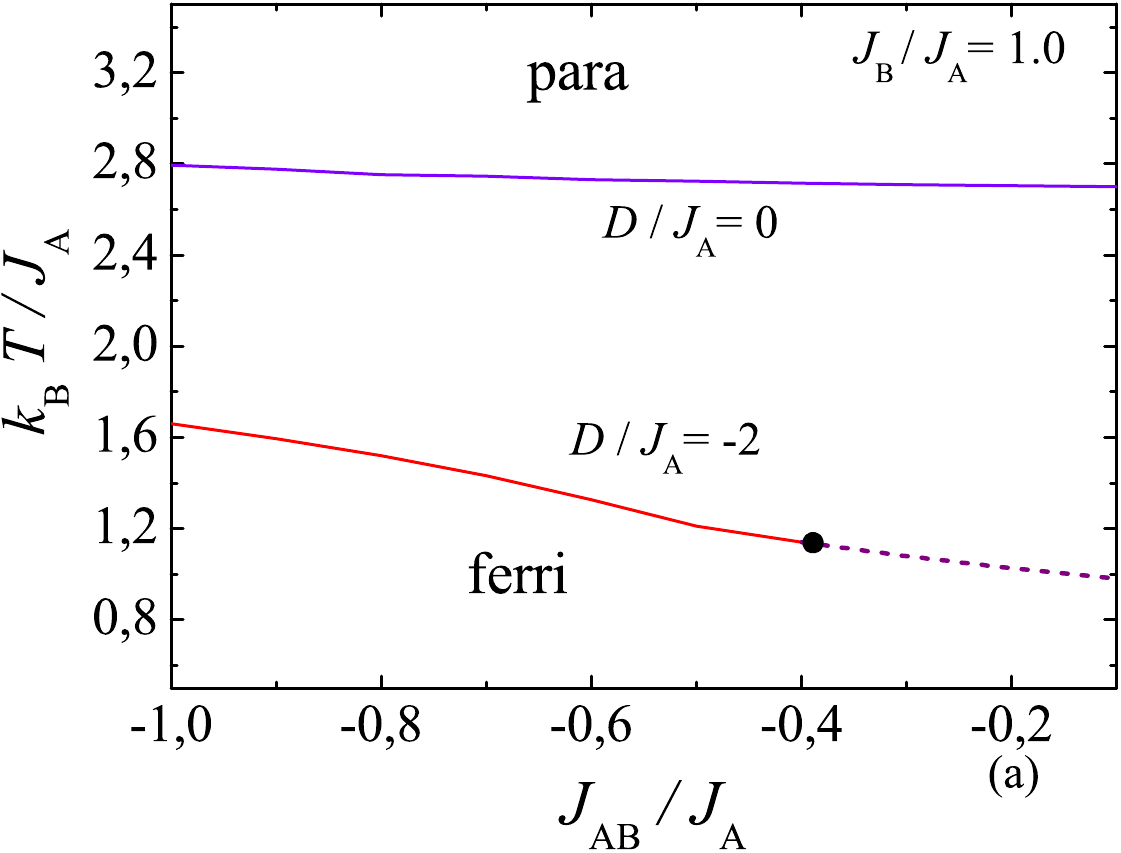}
\includegraphics[width=0.48\textwidth]{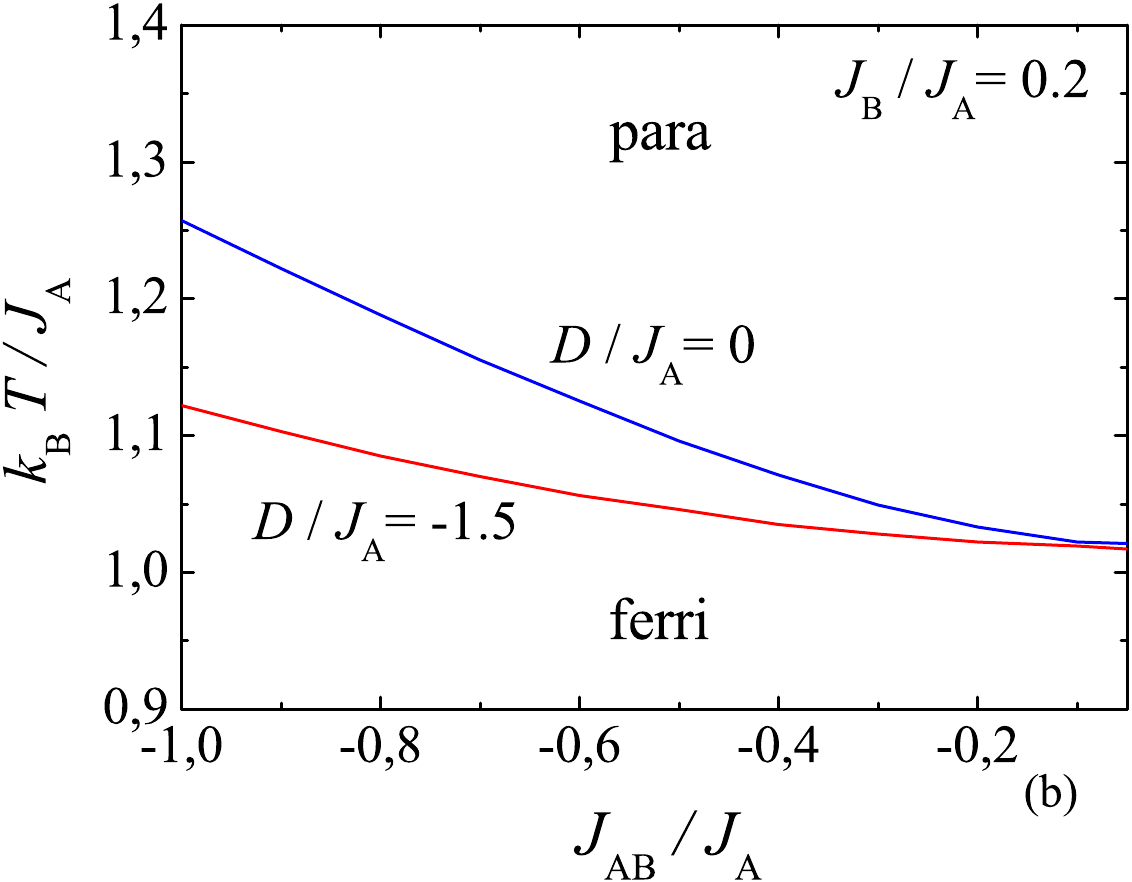}
\caption{(Colour online) The transition temperature as a function of the interlayer coupling for values of the single-ion anisotropy when $J_{B}/J_{A}=1.0$ and $J_{B}/J_{A}=0.2$, respectively. The black circle represents the tricritical point (TCP).}
\label{fig5}       
\end{figure}

Phase diagrams of the ferrimagnetic bilayer Ising bilayer system can be obtained from the phase transitions of monolayer magnetizations. The solid lines in the phase diagrams show the continuous phase transitions and the dashed lines show the first order phase transition. Figure \ref{fig5}(a) shows the phase diagram for $D/J_{A}=0$ and $-2$ in the case of $J_{B}/J_{A}=1$ where the interaction parameters in monolayers are equal. The point where first- and second order phase transition lines join is the tricritical point (TCP). The phase diagram includes the ferrimagnetic region with stable-metastable states and the disordered paramagnetic phase with only stable state. On the other hand, the first-order phase transition line and TCP disappear in the phase diagram for $J_{B}<J_{A}$, when $D/J_{A}=0, -1.5$ [figure \ref{fig5}(b)].

\begin{figure}[!t]
\includegraphics[width=0.47\textwidth]{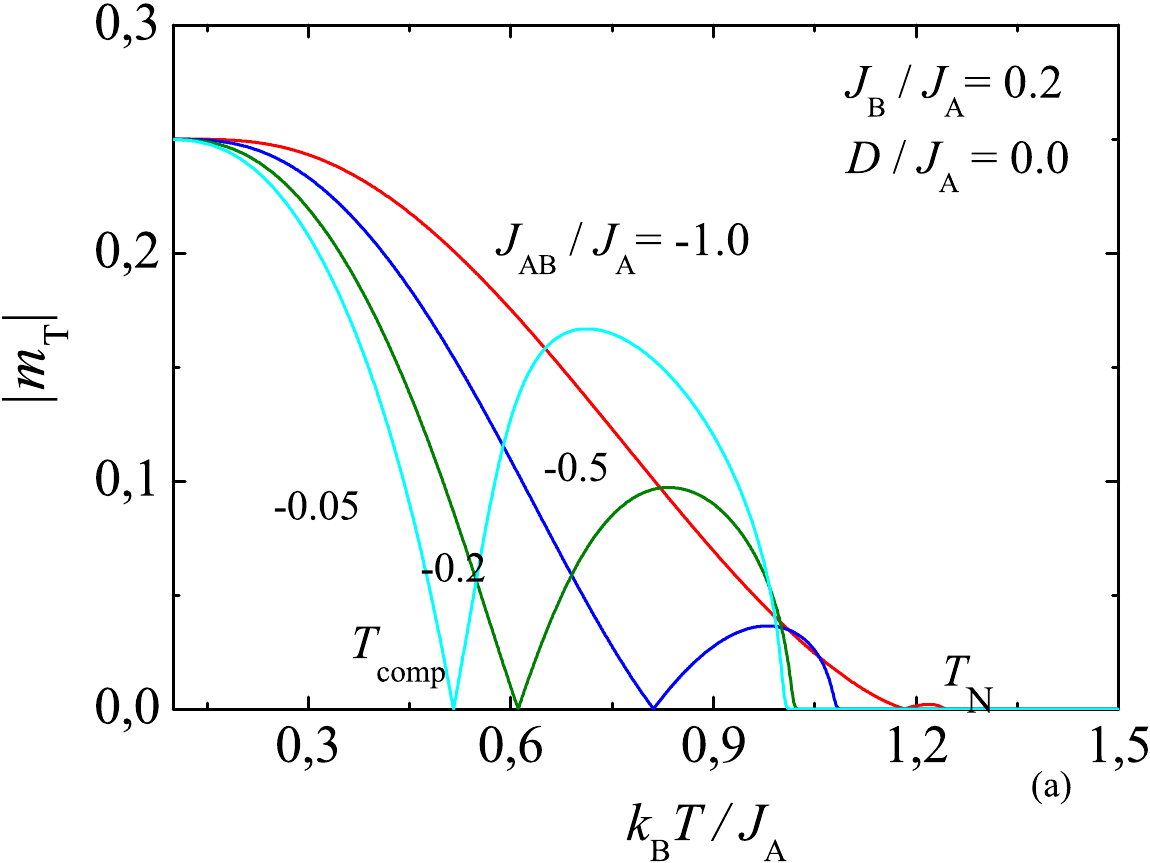}
\includegraphics[width=0.47\textwidth]{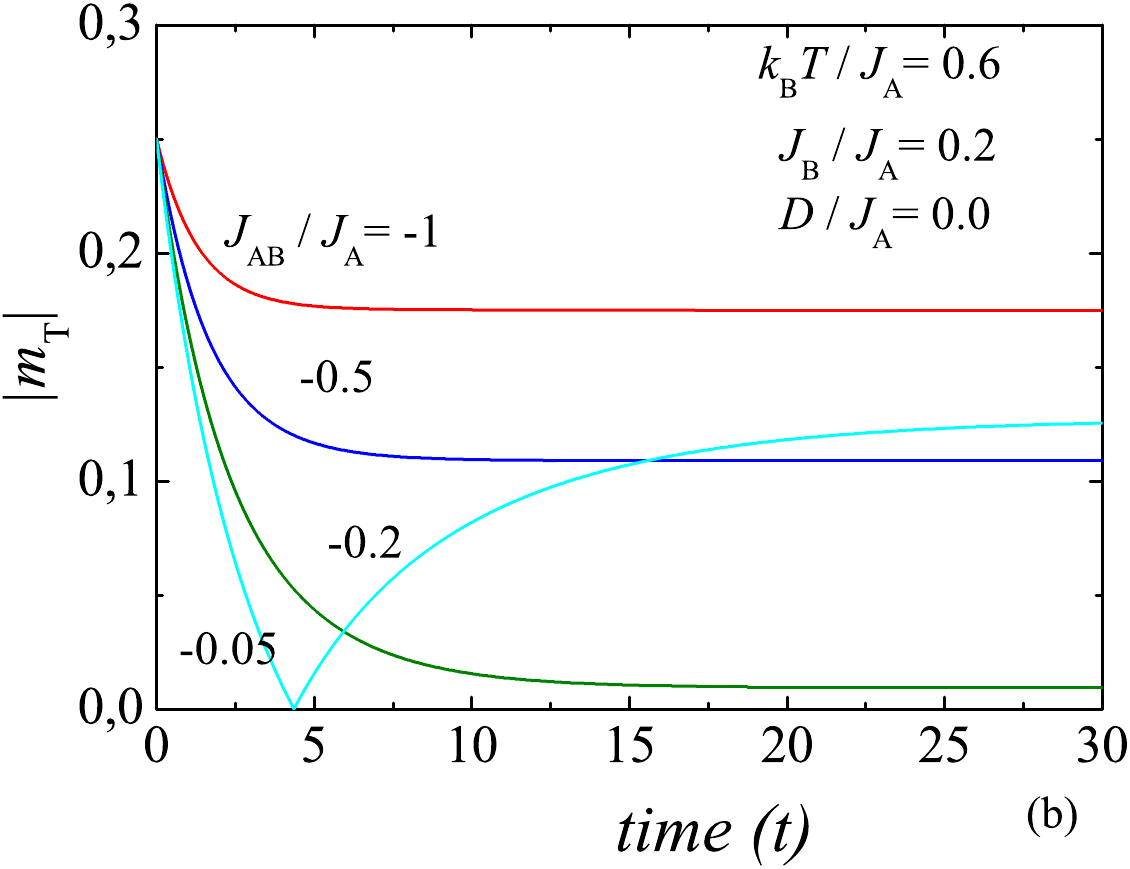}
\caption{(Colour online) (a) The temperature dependence of total magnetization for $J_{B}/J_{A}=0.2$ in the absence of single-ion anisotropy. (b) The time dependence of total magnetization for several interlayer interactions.}
\label{fig6}       
\end{figure}

Next, the equilibrium and dynamic state behaviors of the bilayer Ising model are examined over total magnetization in figure \ref{fig6}. In figure \ref{fig6}(a), the thermal variation of the total magnetization $m_{T}$ is given for the fixed  $J_{B}/J_{A}=0.2$ value for varying antiferromagnetic interlayer interaction parameters $J_{AB}/J_{A} = -1$, $-0.5$, $-0.2$ and $-0.05$. The total magnetization exhibits standard N-type phase transition with compensation temperature according to N\'eel classification for $J_{AB}/J_{A} = -1, -0.5, -0.2, -0.05$. In the bilayer Ising model, which represents a rather simple model of ferrimagnetic materials, the compensation phenomenon is more pronounced for $J_{AB}/J_{A} = -0.05$. At the compensation temperature, the total magnetization $m_{T}$ vanishes, as a result of cancellation between magnetic moments of different monolayers. The significance of compensation temperature is that it can lead to antiferromagnetic-like dynamics in ferrimagnets and high-speed domain walls. Magnetization switching can be easily achieved with faster domain walls, which makes these materials a promising candidate for high-density data storage. 
In figure \ref{fig6}(b), the time evolution of total magnetization is illustrated at $k_{\text{B}}T/J_{A}=0.6$. As can be seen from the time evolution of total magnetizations, all relaxation curves for $J_{AB}/J_{A}= -1, -0.5, -0.2, -0.05$ take a fixed magnetization value i.e., time independent stable state  shown in equilibrium behavior. 
  
\section{Conclusion}
\label{conclusion}
In this study, the bilayer Ising model with different spin magnitudes ($\sigma=1/2$ and $S =1$) for the monolayers with antiferromagnetically coupled interlayer parameter was examined using the lowest approximation of cluster variation method and path probability method. Within the framework of these methods, the dynamic equations giving the variation of monolayer magnetizations with respect to time and order parameters were obtained. Then, the numerical solution of the dynamical equations was performed by the fourth-order Runge-Kutta method and the variations of magnetizations with respect to time were examined in the case of antiferromagnetic interaction between A and B monolayers. In addition to the stable states, the existence of metastable state in the system has been revealed by solving dynamic equations.

Then, the magnetization relations obtained from the dynamical equations of the model were solved and the behavior of the monolayer and total magnetization with respect to temperature was investigated. From the obtained thermal variations, it is seen that for different values of the interaction parameters, both first-order and second-order phase transformations occur in the system. As with the layer structure with different spin magnitudes in previous studies, the phase transition temperatures of the magnetization are the same, but their value varies significantly with temperature depending on the fact that ferromagnetic interaction parameters in each monolayer with different spin values are equal and different. In addition, an increase in the antiferromagnetic interaction parameter between the monolayers reduces the second-order phase transition temperatures and the N\'eel temperature. In the bilayer model, another factor that significantly affects the magnetic behavior of system is the single-ion anisotropy parameter that affects directly the spin-1 states in the monolayer B. The single-ion anisotropy parameter is especially effective on the phase transition in the bilayer system, that is, the first-order phase transition and the compensation temperature. In the variation of total magnetization with respect to temperature at the compensation temperature, the total magnetization first disappears below the second-order phase transition temperature, and this temperature is of great importance in experimental systems, especially in magnetic recording system. Thus, the different spin magnitudes in the monolayers and inclusion of single-ion anisotropy parameter in the system lead to richer phase transitions and the magnetic behavior such as compensation temperature and metastable state.  

\section*{Acknowledgements}
This work was supported by ADU project FEF-11028.

\newpage
\ukrainianpart

\title{Динамічна поведінка ізінгової моделі з антиферомагнітно зв'язаним подвійним шаром}
\author{К. Екіз\refaddr{label1}, Р. Ердем\refaddr{label2}, Д. Семет\refaddr{label3}}
\addresses{
	\addr{label1} Кафедра фізики, природничий факультет, Університет Аднана Мендереса, 09010 Айдин, Туреччина \\ 
	\addr{label2} Кафедра фізики, природничий факультет, Університет Акденіз, 07058 Анталія, Туреччина \\
	\addr{label3} Науково-прикладна лікарня, Університет Аднана Мендереса, 09010 Айдин, Туреччина \\
}

\makeukrtitle

\begin{abstract}
\tolerance=3000%
Використовуючи ймовірності за траєкторіями та найнижче наближення методу варіації кластера, ми досліджуємо динамічні та рівноважні властивості двошарової магнітної системи, що складається з двох феро\-магнiтних моношарів, антиферомагнітно з’єднаних для різних спінів $(\sigma=1/2$ і $S=1) $. Спочатку представлені числові результати моношарової та повної намагніченостей під впливом різних фізичних параметрів, потім обговорюються фазові діаграми двошарової системи. Потім, оскільки добре встановлено, що метод ймовірності за траєкторіями є ефективним для дослідження існування метастабільних станів, вивчається часова еволюція моношарової та повної намагніченостей.
\keywords двошарова модель Ізінга, динамічні властивості, феримагнетизм,  метод ймовірності за траєкторіями

\end{abstract}

\end{document}